\documentclass[aps,prc,twocolumn,showpacs,preprintnumbers,amsmath,amssymb]{revtex4}

\usepackage{amsmath}
\usepackage{graphicx}
\usepackage{dcolumn}
\usepackage{bm}

\newcommand{\JPSI}{$J/\psi$~}
\newcommand{\CHIC}{$\chi_{\rm c}$~}
\newcommand{\PSI}{$\psi^{\prime}$~}
\newcommand{\SNN}{$\sqrt{s_{_{\rm NN}}}$~}
\newcommand{\RAA}{$R_{AA}$~}

\newcommand{\TC}{$T_{\rm c}$~}

\newcommand{\PT}{$p_{_{\rm T}}$~}
\newcommand{\ccbar}{$c\bar{c}$~}

\begin{document}

\preprint{TKYNT-07-06/CNS-REP-70}

\title{Onset of \JPSI Melting in Quark-Gluon Fluid at RHIC}

\author{T.~Gunji}
\email{gunji@cns.s.u-tokyo.ac.jp}
\author{H.~Hamagaki}
\affiliation{%
  Center for Nuclear Study, Graduate School of Science, the University of Tokyo, 7-3-1 Hongo, Bunkyo-ku, Tokyo, 113-0033, Japan
}%

\author{T.~Hatsuda}
\affiliation{%
Department of Physics, Graduate School of Science, the University of Tokyo, 7-3-1 Hongo, Bunkyo-ku, Tokyo, 113-0033, Japan
}%
\author{T.~Hirano}
\affiliation{%
Department of Physics, Graduate School of Science, the University of Tokyo, 7-3-1 Hongo, Bunkyo-ku, Tokyo, 113-0033, Japan
}%

\date{\today} 

\begin{abstract}
  A strong \JPSI suppression in central Au+Au collisions  
  has been observed by the PHENIX experiment at the Relativistic Heavy Ion Collider (RHIC). 
  We develop a hydro+\JPSI model in which hot quark-gluon matter is described by the 
  full (3+1)-dimensional relativistic hydrodynamics and \JPSI is treated
  as an impurity
  traversing through the matter.  The experimental
  \JPSI suppression pattern in mid-rapidity is
  reproduced well by the sequential melting of $\chi_{\rm c}$, $\psi'$, 
  and \JPSI in dynamically expanding fluid.
  The melting temperature of directly produced \JPSI 
  is well constrained by the participant-number dependence of
  the \JPSI suppression and
  is found to be about $ 2 T_{\rm c}$  with $T_{\rm c}$ being the 
  pseudo-critical temperature.
\end{abstract}


\maketitle

\section{Introduction}

A new state of matter composed of
deconfined quarks and gluons, the Quark-Gluon Plasma (QGP),
is expected to be formed 
in relativistic heavy-ion collisions if the system reaches a 
temperature larger than the critical value
$T_{\rm c}\sim160-190$~MeV as predicted by
the  lattice Quantum Chromodynamics (QCD)~\cite{bib:Lat_qgp}.
To find the experimental evidence of QGP, 
the heavy quarkonia ($J/\psi$, $\psi^{\prime}$, $\chi_{\rm c}$, and $\Upsilon$) have
long been considered as the most promising probe, since  
they are supposed to melt away due to the color Debye 
screening at sufficiently high temperature~\cite{bib:Matsui}. 
Recent lattice QCD studies show that
\JPSI would survive even up to  about $2 T_{\rm c}$ while
\CHIC and \PSI would be dissociated at different temperature
~\cite{bib:lattice1,bib:lattice2,bib:lattice3}. 
 Therefore,  the heavy quarkonia may be used as a thermometer of QGP 
in relativistic heavy-ion collision experiments~\cite{bib:Satz}.

Recently, high statistics data of Au+Au collisions at the center 
of mass energy per nucleon ($\sqrt{s_{\rm NN}}$) of 200 GeV at 
the Relativistic Heavy Ion Collier (RHIC) 
in Brookhaven National Laboratory (BNL) become
available~\cite{bib:JPSI_RHIC}.
It is observed that \JPSI yield in central Au+Au collisions at RHIC 
is suppressed by a factor of 4 at mid-rapidity and 5 at forward rapidity 
relative to that in $p+p$ collisions scaled 
by the average number of binary collisions. 
Cold nuclear matter (CNM) effects
due to the gluon shadowing and nuclear absorption 
of \JPSI at the RHIC energy were evaluated from the \JPSI measurement in d+Au
collisions~\cite{bib:JPSI_dAu}. 
Trend of the \JPSI suppression in d+Au collisions as a function of rapidity
is reasonably well reproduced by 
the gluon shadowing with the Eskola-Kolhinen-Salgado (EKS) 
parameterization~\cite{bib:EKS} and  
the nuclear absorption cross section
$\sigma_{\mathrm{abs}} \sim$ 1~mb~\cite{bib:JPSI_dAu}.
Then, a comparison of the \JPSI yield observed in Au+Au collisions at RHIC to that expected from 
CNM effects~\cite{bib:NNS_AuAu} reveals that \JPSI is anomalously suppressed
in central collisions~\cite{bib:QM}. 
 
So far, two scenarios have been proposed for
the \JPSI suppression at RHIC energies.
The first scenario is based on the sequential
melting of the charmonia~\cite{bib:sequential2}: 
The \JPSI suppression previously observed at Super Proton Synchrotron (SPS)
in European Center for Nuclear Research (CERN)
may be ascribed to the 
complete melting of \CHIC and \PSI and associated  
absence of the feed down to $J/\psi$, while
at RHIC, extra melting of \JPSI
could be seen due to higher initial temperature~\cite{bib:chau}.
This approach is still in a qualitative level and the space-time
evolution of the system has not been taken into account.
The second scenario is based on the substantial charmonium dissociation
due to gluons and comovers supplemented with the regeneration
of \JPSI due to \ccbar recombination~\cite{bib:nuxu1,bib:Rapp,bib:thews,bib:anton}.   
In this scenario, a large amount of regeneration is expected at RHIC
in comparison to SPS.

The purpose of this paper is to make a first attempt to investigate
the sequential charmonia suppression in a dynamically evolving matter
produced in Au+Au collisions at RHIC energy.
We develop a hydro+\JPSI model in which hot quark-gluon matter is described by the 
full (3+1)-dimensional relativistic hydrodynamics~\cite{Hirano:2001eu}
and $J/\psi$, $\chi_{\rm c}$, and \PSI are treated as impurities
  traversing through the matter. 
We will focus on the \JPSI data in the mid-rapidity region
throughout this paper, since the hydrodynamical
description of various observables is best established
in that region. The \JPSI suppression in forward rapidity
will be remarked at the end of this paper.

Hereafter we study  the survival probability of $J/\psi$, 
$S^{\mathrm{tot}}_{J/\psi}=R_{AA}/{R_{AA}^{\rm CNM}}$, as a function of the 
number of participants $N_{\rm part}$~\cite{bib:QM}.
\RAA is the standard nuclear modification factor defined by
\begin{equation}
  R_{AA} = \frac{dN^{J/\psi}_{\mathrm{Au+Au}}/{dy}}{\langle N_{\rm col}\rangle dN^{J/\psi}_{p+p}/{dy}},
\label{eq:R_AA}
\end{equation} 
where $dN^{J/\psi}_{\mathrm{Au+Au}}/{dy}$ and $dN^{J/\psi}_{p+p}/{dy}$ are 
the invariant \JPSI yield in Au+Au and $p+p$ collisions, respectively.
$\langle N_{\rm col} \rangle$ is the  average number of nucleon-nucleon collisions.
$R_{AA}^{\rm CNM}$ is a contribution
to \RAA originating from the CNM effects constrained by the data 
of the d+Au collisions.
 
\section{Hydro+$J/\psi$ Model}
\subsection{Full 3D Relativistic Hydrodynamics}
 We solve the equations of energy-momentum conservation
$\partial_\mu T^{\mu\nu} = 0$ in full (3+1)-dimensional space-time
 $(\tau, x, y, \eta_s)$
  under the assumption that
 the local thermal equilibrium is reached 
 at initial time $\tau_0$(=0.6 fm/$c$) \cite{tau0}.
 Here $\tau$ and $\eta_s$ are the proper time and the space-time rapidity, 
 respectively.
 $x$ and $y$ are transverse coordinates. The centers of two colliding nuclei 
 are located at $(x,y) = (b/2,0)$ and $(-b/2,0)$  before collision
  with an impact parameter $b$. The ideal hydrodynamics is
   characterized by the energy-momentum tensor,
 \begin{equation}
 T^{\mu\nu} = (e + P)u^\mu u^\nu - Pg^{\mu\nu}, 
 \end{equation}
where $e$, $P$, and $u^{\mu}$ are 
energy density, pressure, and local four velocity, respectively.
We neglect the finite net-baryon density which is small near the 
 mid-rapidity at RHIC.
 The equation of state (EOS) of  massless parton gas composed of
  $u$, $d$, $s$ quarks and gluons is employed in the QGP phase
 ($T > T_{\rm c}$=170 MeV), while a hadronic resonance gas model is
 employed for $T < T_{\rm c}$.
 The initial energy density distribution in the transverse plane
is parameterized in proportion to $d^{2}N_{\mathrm{col}}/dxdy$
 according to the Glauber model. 
 The pseudo-rapidity distribution of charged particle multiplicity
 $dN^{\mathrm{ch}}/d\eta$ for various centralities observed at RHIC
 has been well reproduced by the (3+1)-dimensional
 hydrodynamics simulations with the above setups~\cite{Hirano:2001eu}.

\begin{figure}[htbp]
  \centering
  \includegraphics[width=0.5\textwidth]{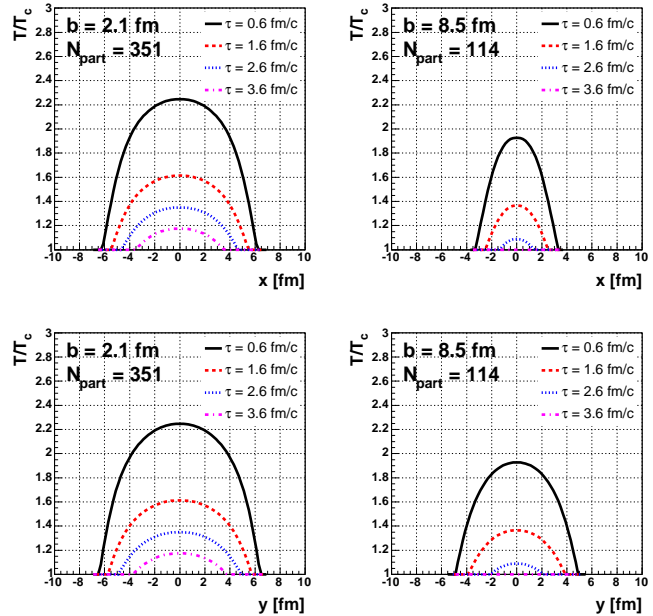}
  \caption{Local temperature in the unit of $T_{\rm c}$
  as functions of the transverse coordinates, 
  $x$ (upper) and $y$ (lower), 
    for various proper time ($\tau$) at mid-rapidity ($\eta_s=0$). The impact parameter
    $b=2.1$~fm (the left panel) corresponds to  $\langle N_{\mathrm{part}} \rangle = 351$,
     while $b=8.5$~fm~(the right panel) corresponds to
      $\langle N_{\mathrm{part}} \rangle = 114$.   
      }
  \label{fig:hydro_T}
\end{figure}

The same space-time evolution of the QGP fluid obtained as above,
 which is now open to public~\cite{parevo},
 has been also exploited to study hard probes
 such as azimuthal jet
 anisotropy,  nuclear modification factor of identified hadrons, and
 disappearance of back-to-back jet correlation~\cite{Hirano:2003yp}.
 Figure~\ref{fig:hydro_T} shows the local temperature as functions of 
$x$ (upper) and $y$ (lower) for various proper time at mid-rapidity ($\eta_s=0$).
The left and right panels 
are for the case at $b=2.1$~fm ($N_{\mathrm{part}}= 351$)
and $b=8.5$~fm~($N_{\mathrm{part}} = 114$), 
respectively.

\subsection{$J/\psi$ inside QGP}

We distribute the initial $J/\psi$'s in the transverse plane at $\eta_s = 0$
according to the spatial distribution of $N_{\mathrm{col}}$ calculated from the
Glauber model for a given impact parameter. Since we focus on the 
mid-rapidity, only transversal motion of \JPSI is considered.
Transverse momentum ($p_{\rm T}$) of \JPSI is distributed according to the 
invariant \PT spectrum of \JPSI measured by PHENIX~\cite{bib:JPSI_RHIC}. 
 This corresponds to a weak coupling case where 
  $J/\psi$ is assumed to follow the free-streaming path
 inside the hot matter unless it is dissociated \cite{free}.

The survival probability of \JPSI  suffering
from  dissociation along its path inside the expanding fluid may be 
expressed as
\begin{equation}
\! \! \! S_{J/\psi}(\bm{x}_{J/\psi}(\tau)) 
= \exp \left[ - \int_{\tau_0}^{\tau} \Gamma_{\rm dis}(T({\bm{x}}_{J/\psi}(\tau')))
d\tau' \right],
\label{eq:SJP}
\end{equation}
where $\Gamma_{\rm dis}(T({\bm{x}}_{J/\psi}(\tau)))$ is the decay width at
temperature $T$ and ${\bm{x}}_{J/\psi}(\tau)$ 
is the transverse
position of \JPSI at proper time $\tau$.
Information on $\Gamma_{\rm dis}(T)$ 
is rather scarce at present because the matter is still
in non-perturbative region $T/T_{\rm c} \sim 1 - 2$. 
 We will first make a simplest threshold ansatz according to
 the similar one in \cite{bib:BO}:
 $\Gamma_{\rm dis}(T > T_{J/\psi}) = \infty$
 and $\Gamma_{\rm dis}(T < T_{J/\psi}) = 0$
 where $T_{J/\psi}$ is the melting temperature of $J/\psi$. 
 A more sophisticated parametrization of 
 $\Gamma_{\rm dis}(T)$ will be discussed later.

 We assume the same \PT distribution for 
 $\chi_{\rm c}$ and $\psi '$ as that for \JPSI and 
  estimate their survival probabilities in a
similar way as Eq.~(\ref{eq:SJP})
 with a common melting temperature
$T_{\psi'}=T_{\chi_{\rm c}}\equiv T_{\chi}$.
Thus, the total survival probability of \JPSI is obtained by taking into
account the total  feed down fraction  $f_{_{\rm FD}}$ from \CHIC + \PSI
to \JPSI as 
\begin{equation}
  \label{eq:seq_toy}
  S^{\rm tot}_{J/\psi} = (1-f_{_{\rm FD}})S_{J/\psi} + f_{_{\rm FD}}S_{\chi},
\end{equation}
where $S_{\chi}$ denotes the  survival probability of \CHIC or \PSI.

Although the precise values of 
$T_{J/\psi,\chi_{\rm c},\psi'}$ should
 be eventually calculated from lattice QCD simulations with dynamical light quarks,
 we take those as free parameters to fit the present experimental 
 data of $S^{\rm tot}_{J/\psi}$.
 The feed down fraction $f_{_{\rm FD}}$ is also treated as a free parameter.
  At present, its value measured at different energies has a large uncertainty,
   $15 \% - 74 \%$~\cite{bib:FD}. Also it 
  may well depend on collision energy and has not yet been measured  at the RHIC energy.
As we will see below, the shape (magnitude) of $S^{\rm tot}_{J/\psi}$ as a 
function of $N_{\mathrm{part}}$ is essentially determined by
 $T_{J/\psi}$ ($f_{_{\rm FD}}$).

\subsection{Numerical Results}
\begin{figure}[bthp]
  \centering
  \includegraphics[width=0.47\textwidth]{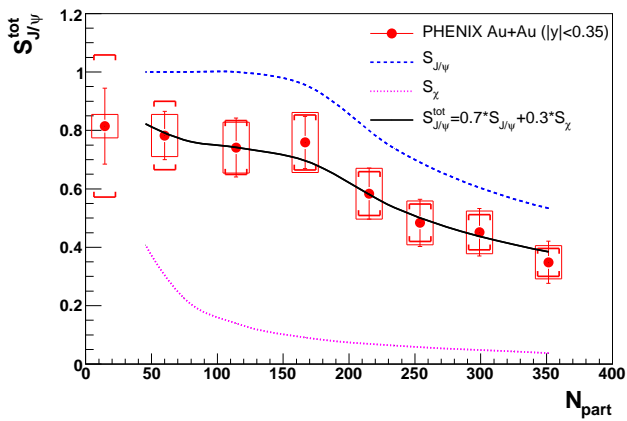}
  \caption{Survival probabilities $S^{\rm tot}_{J/\psi}$ (solid line),
   $S_{J/\psi}$ (dashed line), 
    and $S_{\chi}$ (dotted line) in the hydro+\JPSI model
     as a function of the number of participants
     $N_{\mathrm{part}}$
    with $(T_{J/\psi}, T_{\chi}, f_{_{\rm FD}})=(2.02T_{c},1.22T_{c}, 0.30)$.
    Filled symbols are the experimental 
    survival probability in the mid-rapidity of \JPSI in Au+Au collisions at RHIC~\cite{bib:JPSI_RHIC, bib:QM}.
        }
  \label{fig:survival_prob_Jpsi_diss_21_12}
\end{figure}

 Experimental survival probability
  of \JPSI  in the mid-rapidity region in Au+Au collisions at RHIC
 is shown as filled circles in Fig.~\ref{fig:survival_prob_Jpsi_diss_21_12}, 
 where gluon shadowing and nuclear absorption with $\sigma_{\rm abs}=$ 1~mb are
 taken into account as CNM effects~\cite{bib:QM}. 
 Bars and brackets correspond to the uncorrelated and correlated 
 errors with respect to $N_{\rm part}$, respectively~\cite{bib:JPSI_RHIC}.
 Boxes correspond to the uncertainties associated with the 
 nuclear absorption cross section of 0~mb $-$ 2~mb~\cite{bib:QM}.

 The solid line corresponds to the net survival probability 
 $S^{\rm tot}_{J/\psi}$ obtained from our hydro+\JPSI model
 with the best fit parameters ($\chi^{2}=0.86$ for $N_{\mathrm{part}} \ge 50$), 
\begin{eqnarray}
(T_{J/\psi}, T_{\chi}, f_{_{\rm FD}})=(2.02T_{\rm c},1.22T_{\rm c}, 0.30),
\label{eq:best_fit}
\end{eqnarray}
while $S_{J/\psi}$ and $S_{\chi}$ are shown by dashed and dotted lines, respectively.

Decreasing of $S^{\rm tot}_{J/\psi}$ with increasing
$N_{\mathrm{part}}$ is reproduced quite well with the scenario of
 sequential melting in expanding fluid: 
Onset of genuine \JPSI suppression 
around $N_{\mathrm{part}} \sim$ 160 can be clearly seen by the 
 dashed line in Fig.~\ref{fig:survival_prob_Jpsi_diss_21_12}, which 
 results from the
fact that the highest temperature of the matter 
reaches to $T_{J/\psi}$ at this centrality 
(See Fig.~\ref{fig:hydro_T} for comparison).
 Gradual decrease of $S^{\rm tot}_{J/\psi}$ above $N_{\mathrm{part}} \sim$ 160 
 reflects the fact that the transverse area with  $T > T_{J/\psi}$ 
 increases.

\begin{figure}[bthp]
  \centering
 \includegraphics[width=0.47\textwidth]{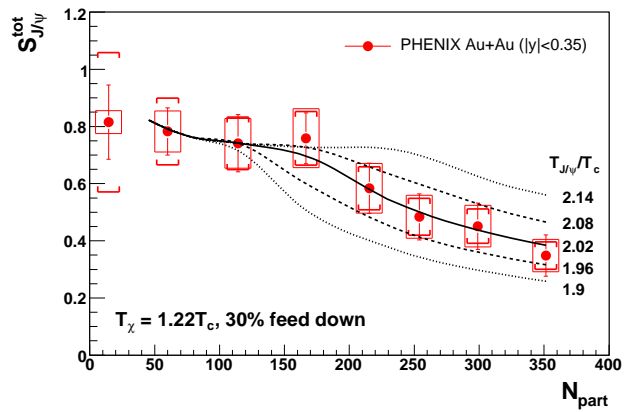}
  \caption{Survival probability $S^{\rm tot}_{J/\psi}$ as a function of 
    the number of participants $N_{\mathrm{part}}$, where melting temperature of \JPSI is 
    1.90\TC (dashed), 1.96\TC (dotted), 2.02\TC (solid with open circles), 
    2.08\TC (dash-dotted), and 2.14\TC (solid).}
  \label{fig:saa_Tdiss_JPsi}
\end{figure}

\begin{figure}[tbhp]
  \centering
  \includegraphics[width=0.5\textwidth]{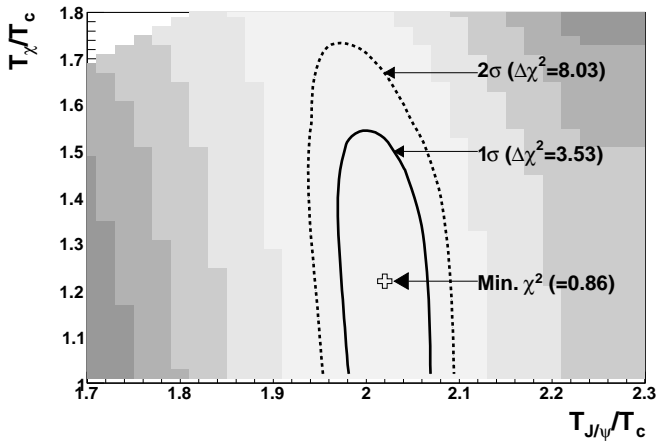}
  \caption{$\chi^{2}$ contour plot in $T_{J/\psi}/T_{\rm c}-T_{\chi}/T_{\rm c}$
   plane  with $f_{_{\rm FD}}$ being fixed to 0.30.
   Cross symbol corresponds to $(T_{J/\psi}, T_{\chi})=(2.02T_{\rm c},1.22T_{\rm c})$, which gives
    minimum $\chi^{2}$. Solid and dashed lines correspond to the $1\sigma$ and $2\sigma$              
    contours, respectively. }
  \label{fig:cont}
\end{figure}

Sensitivity of $S^{\rm tot}_{J/\psi}$ to
the melting temperature of \JPSI is shown 
in Fig.~\ref{fig:saa_Tdiss_JPsi}. 
Here, $f_{_{\rm FD}}$ and $T_{\chi}$
  are  fixed to be 30\% and 1.22$T_{\rm c}$, respectively, as given 
in Eq.~(\ref{eq:best_fit}).
The shape of the 
theoretical curve of $S^{\rm tot}_{J/\psi}$ is  sensitive to
$T_{J/\psi}$:  Even the 6\% change of the 
melting temperature of \JPSI from the best fit
value 2.02 \TC causes a substantial 
deviation from  the experimental data.
To make this point clearer,
we show, in Fig.~\ref{fig:cont}, the $\chi^{2}$ contour  plot in the plane of 
$T_{J/\psi}/T_{\rm c}$ and $T_{\chi}/T_{\rm c}$
with $f_{_{\rm FD}}$ being fixed to 0.30. 
The cross symbol corresponds to Eq.~(\ref{eq:best_fit})
which gives  minimum $\chi^{2}$. 
Solid and dashed lines correspond to the $1\sigma$ and $2\sigma$ 
contours, respectively.
It is seen that $T_{J/\psi}$ can be determined 
in  a narrow region around  $2 T_{\rm c}$, while
$T_{\chi}$ is not well determined 
because the feed-down effect is not a dominant contribution
to $S^{\rm tot}_{J/\psi}$.

 $T_{J/\psi}$ is also insensitive to the nuclear absorption cross section.
  Taking  $\sigma_{\rm abs} =$ 0~mb (2~mb) as an upper (lower)
  bound~\cite{bib:QM},  we found 
  $(T_{J/\psi}, T_{\chi}, f_{_{\rm FD}})=(2.00T_{\rm c},1.02T_{\rm c}, 0.35)$ 
  with minimum $\chi^{2}$ of 0.97 for $N_{\mathrm{part}} \ge 50$ 
  ($(T_{J/\psi}, T_{\chi}, f_{_{\rm FD}})=(2.02T_{\rm c},1.02T_{\rm c}, 0.15)$ 
  with minimum $\chi^{2}$ of 0.87 for $N_{\mathrm{part}} \ge 50$).
  The change of  $\sigma_{\rm abs}$ affects only the overall normalization
  of  $S^{\rm tot}_{J/\psi}$, whose effect 
  is  mostly absorbed by the change of $f_{_{\rm FD}}$.
  In other words, the value of $T_{J/\psi}$ is intimately 
   linked to the shape of  $S^{\rm tot}_{J/\psi}$ around
  $N_{\mathrm{part}} \sim$ 160.

It is noticeable that the RHIC data analyzed with the 
state-of-the-art hydrodynamics leads to a rather  stable value
for the melting temperature of \JPSI to be around $T/T_{\rm c} \simeq 2$ ~\cite{bib:lattice_comp}.
In fact, this number is not in contradiction to the 
result of recent quenched and full lattice QCD simulations
which suggest the survival of \JPSI above $T_{\rm c}$~\cite{bib:lattice1,bib:lattice2,bib:lattice3}.
To make a quantitative comparison, however, 
we need to wait for further progresses in lattice QCD
simulations with dynamical quarks \cite{bib:lattice3}.

\subsection{Thermal broadening of $J/\psi$}
\label{sec:tb}
 
The results presented so far are obtained by assuming that the decay width of charmonia 
changes abruptly from zero to infinity across the melting temperature.
 To make smooth connection with the scenario of the 
 $J/\psi$ dissociation due to thermal quarks and gluons in QGP,
 we introduce the following parametrization for the charmonia width
 motivated by the $T$-dependence in  ~\cite{bib:SHL_decay_width}:
\begin{equation}
\Gamma_{\rm dis}(T < T_{J/\psi}) = \alpha(T/T_{\rm c}-1)^2,
\label{eq:dw}
\end{equation}
with $\Gamma_{\rm dis}(T > T_{J/\psi}) = \infty$.
 $\alpha$ is nothing but the thermal width of $J/\psi$ at $T/T_{\rm c}=2$.
 The $J/\psi$ dissociation width in NLO perturbative QCD calculation   
 suggests $\alpha > 0.4 $ GeV  ~\cite{bib:SHL_decay_width}, 
 while we leave it as a free parameter
 to take into account the fact that the system is still in the 
 non-perturbative regime.

\begin{figure}[tbhp]
  \centering
  \includegraphics[width=0.5\textwidth]{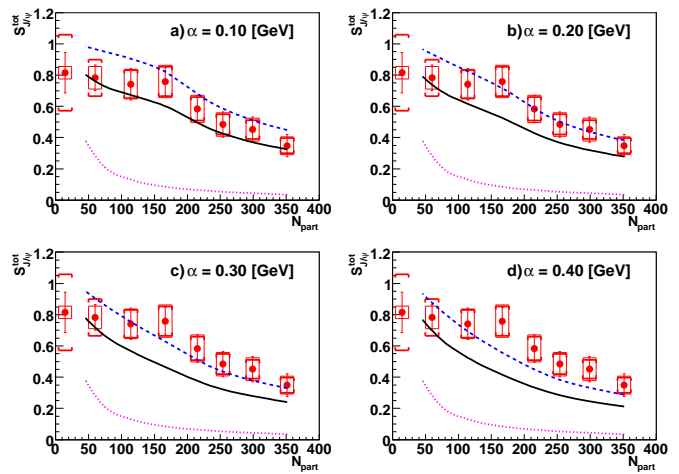}
  \caption{Survival probability $S^{\rm tot}_{J/\psi}$ as a function of 
    $N_{\mathrm{part}}$, where melting temperatures of $J/\psi$ and $\chi$ 
    and feed down fraction are 2.02$T_{\rm c}$, 1.22$T_{\rm c}$ and 30\%.
    a), b), c), and d) are the results using the decay width 
    parameterized by Eq.~(\ref{eq:dw}) with $\alpha$ of 0.1, 0.2, 0.3, and 
    0.4 GeV, respectively.}
  \label{fig:saa_alpha_dep}
\end{figure}
Figure~\ref{fig:saa_alpha_dep} shows the survival probability as a 
function of $N_{\rm part}$, where the model calculations 
shown in a), b), c), and d) are obtained for 
$\alpha = 0.1, 0.2, 0.3$, and 0.4 GeV, respectively.
 The melting temperatures of $J/\psi$ and $\chi$
and feed down fraction are fixed to be 
2.02$T_{\rm c}$, 1.22$T_{\rm c}$ and 30\%.
 It is seen that the suppression pattern has a smooth
 change from the shoulder shape to the monotonic shape
  as $\alpha$ increases. The data are not reproduced well
   for $\alpha > 0.1$ GeV. It is unclear 
   whether the gap between the 
   data and the solid curve for $\alpha > 0.1 $ GeV 
   can be filled by the charmonium regeneration, since
   the regeneration is a monotonically increasing function of $N_{\rm part}$.

\subsection{$J/\psi$ in Forward Rapidity}
 
 Let us here make a brief remark on the \JPSI
 suppression at forward rapidity which is reported to be
 even stronger  than that at mid-rapidity~\cite{bib:JPSI_RHIC}.
 Experimentally, there is still a sizable systematic error 
 for relative normalization between the two cases~\cite{bib:QM}.
 This could reduce the apparent difference  of $R_{AA}$  
 between the data at mid and forward rapidities.
 Theoretically, a part of the difference may come from gluon saturation.
 According to Color-Glass-Condensate (CGC) model,
 the charm production at forward rapidity ($y \sim 2$) may become as low as 
  60\% of that at mid-rapidity in most central Au+Au collisions~\cite{bib:CGC}.
  If we take this number literally, the \JPSI suppressions in both rapidities
  become almost comparable. There is also a possibility that
  $\sigma_{\rm abs}$ depends on the \JPSI rapidity and is larger in 
  forward rapidity~\cite{bib:ABS}. 
  With all these uncertainties, we leave the \JPSI suppression
  in forward rapidity as an interesting subject for future works.
  
\section{Summary}

In summary, we have investigated \JPSI yield at mid-rapidity
 in the relativistic heavy ion collisions by using hydro+\JPSI model which
 serves as a dynamical approach to the sequential charmonia suppression.
 The space-time 
  dependence of the local temperature $T$ of the fluid is described by the 
  state-of-the-art relativistic hydrodynamic simulations, and 
 the \JPSI dissociation is assumed when $T$ exceeds
  the melting temperature $T_{J/\psi}$.
 Comparing  the survival probability $S_{J/\psi}^{\rm tot}$
    from the hydro+\JPSI model with the one 
  in Au+Au collisions at \SNN = 200 GeV obtained 
  by the PHENIX experiment at RHIC,
  observed \JPSI suppression pattern is described quite well 
  with $T_{J/\psi}/T_{\rm c} \simeq 2$. 
  This value is determined primary by the $N_{\rm part}$ dependence
  of $S_{J/\psi}^{\rm tot}$ and is 
  insensitive to the magnitude of the nuclear absorption.
  Occurrence of the suppression of directly produced \JPSI around 2 \TC 
  is in accordance with the spectral analysis of \JPSI in lattice QCD 
  simulations. Trend of the suppression can be described well with 
  the small decay width inside the fluid below the melting temperature. 
 
  There are a number of issues to be studied
  further: They include
   (i) better treatment of the \JPSI propagation in hot matter beyond
    the free-streaming or the complete thermalization,
   (ii) detailed study on the azimuthal anisotropy ($v_2$) of \JPSI,  
   (iii) effect of relative velocity between \JPSI and
   the hot fluid on the $p_{\rm T}$ distribution of \JPSI \cite{bib:hot_wind},
   (iv) better understanding of the 
    suppression in forward rapidity in connection with
    the color glass condensate, 
   (v) the application to lighter system
    such as Cu+Cu one,  
   (vi) effect of the charmonia regeneration
    processes to  $S_{J/\psi}^{\rm tot}$
    in conjunction with the sequential suppression, and 
   (vii) the theoretical investigation of the melting temperatures
   and the widths
   of $J/\psi$, $\chi_{\rm c}$ and $\psi'$ in full QCD simulations.

 The authors would like to thank M. Natsuume and K. Rajagopal for the
 discussions on the $J/\psi$ dissociation under the hot wind.
 T.~Gunji was the JSPS Research Fellow and this work was partially supported 
 by JSPS grant No.16-11332 (T.~Gunji), No.14204021 (H.~Hamagaki), 
 No.18540253 (T.~Hatsuda) and No.18-10104 (T.~Hirano).


\begin{thebibliography}{10}
\bibitem{tau0} 
 The $\tau_0$ is one of the 
 parameters setting the initial condition of the plasma and 
 is conventionally chosen to be 0.6 fm/$c$ in
 hydrodynamic simulations at RHIC energies.
 Sensitivity of the survival probability of $J/\psi$ 
  under the variation of $\tau_0$ is a non-trivial issue
  and must be studied together with the re-tuning
 of other initial parameters.
\bibitem{free} 
 By using the same initial
  distribution of $J/\psi$, we have checked that
 $R_{AA}$ is not affected even if $J/\psi$ is fully thermalized and flow with the 
  background matter. The difference between the free-streaming and complete
  thermalization appears, however, in the  final $p_{\rm T}$ distribution
  of $J/\psi$. Further details will be reported elsewhere.
\bibitem{bib:lattice_comp}
It would be interesting to estimate $T_{J/\psi}$ in 
3-flavor QCD from that obtained in the quenched QCD simulations.
If we assume that the \JPSI melting is related to the 
 energy density of surrounding medium, it is straightforward to
 see that  $\left( T_{J/\psi}/T_{\rm c} \right)_{\rm f}
 = (16/47.5)^{1/4} \cdot \left( T_{J/\psi}/T_{\rm c} \right)_{\rm q}
 \cdot \left( T_{\rm c}^{({\rm q})}/T_{\rm c}^{({\rm f})} \right)$,
 where f stands for full and q stands for quenched.
 Putting the numbers, 
 $\left( T_{J/\psi}/T_{\rm c} \right)_{\rm q} \simeq 1.7$~\cite{bib:lattice1} and 
 $ T_{\rm c}^{({\rm q})}/T_{\rm c}^{({\rm f})} \simeq 270/170$, we 
 obtain $\left( T_{J/\psi}/T_{\rm c} \right)_{\rm f} \simeq 2.1$
 which coincides with the value  in Eq.(\ref{eq:best_fit}).
\end{thebibliography}

\end{document}